\documentclass{article}
\usepackage[utf8]{inputenc}
\usepackage{authblk}
\usepackage{graphicx}
\begin{document}

\title{All-optical ultrafast ReLU function for energy-efficient nanophotonic deep learning}

\author[1,$\dagger$]{Gordon H.Y. Li}
\author[2,$\dagger$]{Ryoto Sekine}
\author[2,$\dagger$]{Rajveer Nehra}
\author[2,$\dagger$]{Robert M. Gray}
\author[2,3]{Luis Ledezma}
\author[2]{Qiushi Guo}
\author[1,2,*]{Alireza Marandi}
\affil[1]{Department of Applied Physics, California Institute of Technology, Pasadena, CA 91125, USA}
\affil[2]{Department of Electrical Engineering, California Institute of Technology, Pasadena, CA 91125, USA}
\affil[3]{Jet Propulsion Laboratory, California Institute of Technology, Pasadena, California 91125, USA}
\affil[$\dagger$]{These authors contributed equally}
\affil[*]{marandi@caltech.edu}

\date{}
\maketitle

\begin{abstract}
In recent years, the computational demands of deep learning applications have necessitated the introduction of energy-efficient hardware accelerators. Optical neural networks are a promising option; however, thus far they have been largely limited by the lack of energy-efficient nonlinear optical functions. Here, we experimentally demonstrate an all-optical Rectified Linear Unit (ReLU), which is the most widely used nonlinear activation function for deep learning, using a periodically-poled thin-film lithium niobate nanophotonic waveguide and achieve ultra-low energies in the regime of femtojoules per activation with near-instantaneous operation. Our results provide a clear and practical path towards truly all-optical, energy-efficient nanophotonic deep learning.
\end{abstract}

\section{Introduction}
Over the past decade, deep learning has revolutionized many important applications including computer vision, speech recognition, and natural language processing~\cite{goodfellow2016deep}. However, the explosive growth of modern deep learning models has quickly outpaced improvements in conventional von Neumann computing architectures and ushered in the use of dedicated hardware accelerators. The quest for ever-faster and more energy-efficient hardware for deep learning began with exploiting the graphics processing unit (GPU), then application-specific integrated circuits such as Google's tensor processing unit (TPU), and more recently the development of non-von Neumann analog architectures~\cite{sze2017efficient,lecun20191}. Naturally, photonics has attracted attention as a promising candidate due to its potential for massive parallelism and ultrafast operation~\cite{wetzstein2020inference}. Indeed, optical neural networks (ONNs) have been experimentally demonstrated in a variety of platforms including free-space optics~\cite{lin2018all,zhou2021large,zuo2019all,wang2021optical,gu2021optronic,miscuglio2020massively,porte2021complete}, optical fiber~\cite{xu202111,mourgias2019all,duport2012all,duport2016fully,shastri2016spike,dejonckheere2014all}, and photonic integrated circuits~\cite{shen2017deep,feldmann2019all,feldmann2021parallel,ashtiani2021single,xu2021optical}.

In general, deep neural networks require two major types of computations: (1) linear operations in the form of matrix multiplications and convolutions, which represent the synaptic connections of the network, and (2) nonlinear activation functions, which represent the neuron activations. ONNs excel at performing energy-efficient linear operations in the optical domain, which forms the bulk of computations for deep learning. However, a major remaining roadblock is achieving scalable energy-efficient nonlinear activation functions, which comprises a smaller but essential part of the deep learning workload. Thus, the majority of ONN implementations still opt to utilize digital electronics to perform the nonlinear activation functions. In doing so, the optoelectronic and analog-to-digital conversion typically imposes significant speed and energy limitations. On the other hand, the demonstrated all-optical approaches based on various processes~\cite{dejonckheere2014all,feldmann2019all,zuo2019all,mourgias2019all,shi2021inp} are still too energy-intensive and/or slow compared to electronics. This is because photon-photon interactions are typically weak and require either high light intensities or high-Q resonant cavities, both of which are undesirable for scalable computing purposes. An all-optical, ultrafast, and energy-efficient nonlinear activation function is yet to be demonstrated to unlock the full capabilities of ONNs. Such a function should also be compact, highly scalable, and compatible with existing deep learning models. 

In this work, we propose and experimentally demonstrate the first photonic device, to the best of our knowledge, that satisfies all the aforementioned criteria for an all-optical nonlinear activation function. It implements the Rectified Linear Unit (ReLU) function, defined as $\mathrm{ReLU}(x)=\mathrm{max}(0,x)$, which is one of the most widely used nonlinear activation functions for deep learning. The widespread adoption of the ReLU function was essential in sparking the deep learning revolution due to its favorable properties for backpropagation training and simple implementation in digital electronics~\cite{goodfellow2016deep}. However, its optical implementation has remained challenging and posed a major hurdle for the real-world applicability of ONNs.

\section{Method}
\subsection{Principle of operation}
\begin{figure}[!t]
\centering
\includegraphics[scale=0.79]{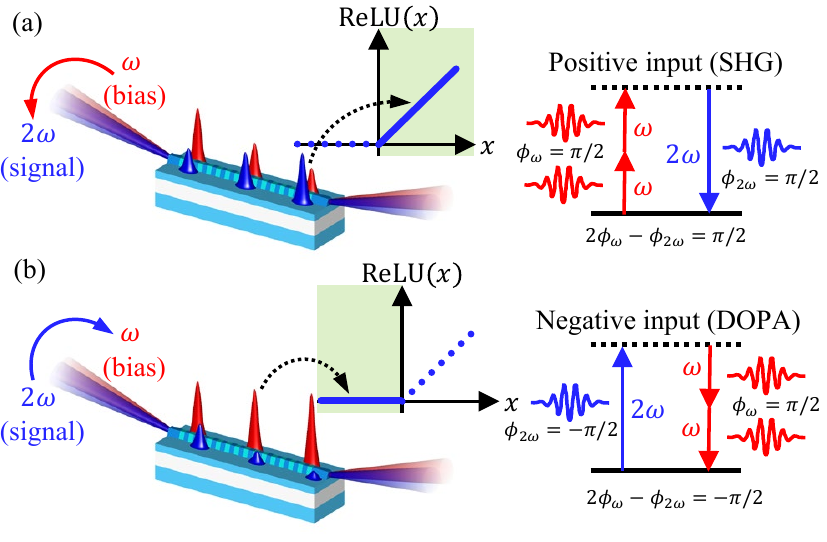}
\caption{Operating principle of the all-optical ReLU function using a nonlinear photonic waveguide. (a) For positive inputs with phase of $\phi_{2\omega}=+\pi/2$, the phase relationship between the signal and bias is $2\phi_{\omega}-\phi_{2\omega}=\pi/2$, which causes SHG that depletes $\omega$ and amplifies $2\omega$. (b) For negative inputs, $\phi_{2\omega}=-\pi/2$, the phase relationship $2\phi_{\omega}-\phi_{2\omega}=3\pi/2\rightarrow-\pi/2$ causes DOPA that amplifies $\omega$ and depletes $2\omega$.}
\label{fig:fig1}
\end{figure}

The operating principle of our device is illustrated in Fig.~\ref{fig:fig1}.
We encode the signal information into the coherent optical field of pulses centered at frequency $2\omega$, with positive values represented by $\phi_{2\omega}=+\pi/2$ phase states, and negative values represented by $\phi_{2\omega}=-\pi/2$ phase states. By co-propagating the signal pulses with bias pulses centered at frequency $\omega$, with fixed input power and phase at $\phi_{\omega}=+\pi/2$, we can induce different nonlinear optical effects for the two possible $\phi_{2\omega}$ signal phases depending on the value of the phase relationship $2\phi_{\omega}-\phi_{2\omega}$. For the positive signal values with phase $\phi_{2\omega}=+\pi/2$, the phase relationship yields $2\phi_{\omega}-\phi_{2\omega}=+\pi/2$. This induces second harmonic generation (SHG), which is a $\chi^{(2)}$ nonlinear optical process that converts two photons of frequency $\omega$ into a photon of frequency $2\omega$, hence depleting $\omega$ and amplifying $2\omega$. Conversely, for the negative signal values with phase $\phi_{2\omega}=-\pi/2$, the phase relationship yields $2\phi_{\omega}-\phi_{2\omega}=3\pi/2\rightarrow -\pi/2$. This induces degenerate optical parametric amplification (DOPA), which is the inverse process of SHG that converts a photon of frequency $2\omega$ into two photons of frequency $\omega$, hence depleting $2\omega$ and amplifying $\omega$. By judiciously choosing the length and bias power, we can achieve the desired shape of the ReLU function. We emphasize that our approach utilizes coherent parametric processes which allows us to implement both positive and negative values, unlike previous optical and optoelectronic methods based on incoherent absorption processes that can only implement positive values.

\subsection{Device design}
\begin{figure}[!h]
\centering
\includegraphics[scale=0.79]{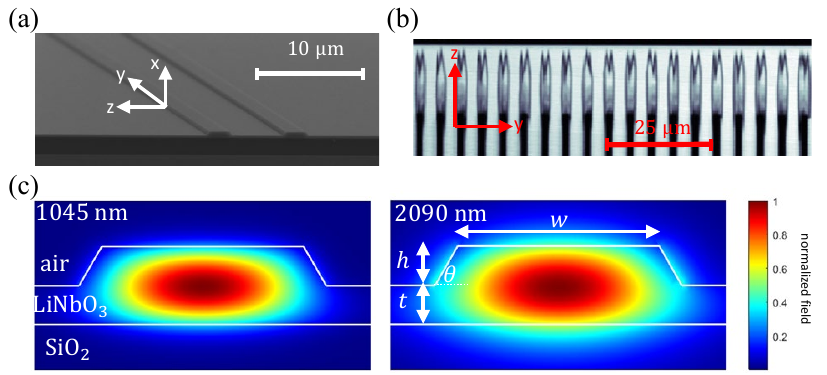}
\caption{Images of the PPLN nanophotonic waveguide. (a) Scanning electron microscope image of the ridge waveguide. (b) Two-photon absorption microscope image of the PPLN ferroelectric domains with poling period of 5$~\mathrm{\mu}$m. (c) Simulated electric field distributions of the fundamental TE modes at $1045~\mathrm{nm}$ ($2\omega$) and $2090~\mathrm{nm}$ ($\omega$).}
\label{fig:fig2}
\end{figure}

To implement the $\chi^{(2)}$-based ReLU function, we use a periodically poled thin-film lithium niobate (PPLN) nanophotonic waveguide that exploits the strong and instantaneous $\chi^{(2)}$ optical nonlinearity of lithium niobate and tight spatial confinement of the waveguide modes to enhance the nonlinearity~\cite{wang2018ultrahigh}. Additionally, careful qausi-phase matching and dispersion engineering enables ultra-broadband and low-energy interactions over mm-long propagation lengths, further enhancing the nonlinear optical processes using femtosecond laser pulses~\cite{jankowski2020ultrabroadband,guo2021femtojoule,ledezma2021intense}. Images of the device are shown in Fig.~\ref{fig:fig2}. The PPLN nanophotonic waveguide is $L=2.5~\mathrm{mm}$ long and was fabricated on a $700$-nm thick X-cut MgO-doped lithium niobate thin-film on 2-$\mathrm{\mu}$m thick SiO$_2$ with lithium niobate substrate by dry etching with $\mathrm{Ar}^{+}$ plasma, achieving smooth ridge side-walls with slant angle of $\theta\approx 60^{\circ}$ as shown in Fig.~\ref{fig:fig2}(a). The waveguide was electrically poled with a period of 5.17$~\mathrm{\mu}$m, as shown in Fig.~\ref{fig:fig2}(b), to ensure efficient SHG and DOPA. Dispersion engineering of the fundamental TE mode of the ridge waveguide, shown in Fig.~\ref{fig:fig2}(c), allows for negligible group velocity mismatch and group velocity dispersion of $\omega$ and $2\omega$ pulses centered at $1045~\mathrm{nm}$ and $2090~\mathrm{nm}$, respectively. This enforces good temporal overlap of the pulses over the entire PPLN propagation length. The ideal parameters found from simulation were a ridge top width of $w=1700~\mathrm{nm}$ and etch-depth of $h=350~\mathrm{nm}$. See~\cite{ledezma2021intense} for further details about fabrication and dispersion engineering of PPLN nanophotonic waveguides.

\section{Results}
\subsection{Femtojoule ReLU function}
\begin{figure}[!t]
\centering
\includegraphics[scale=0.79]{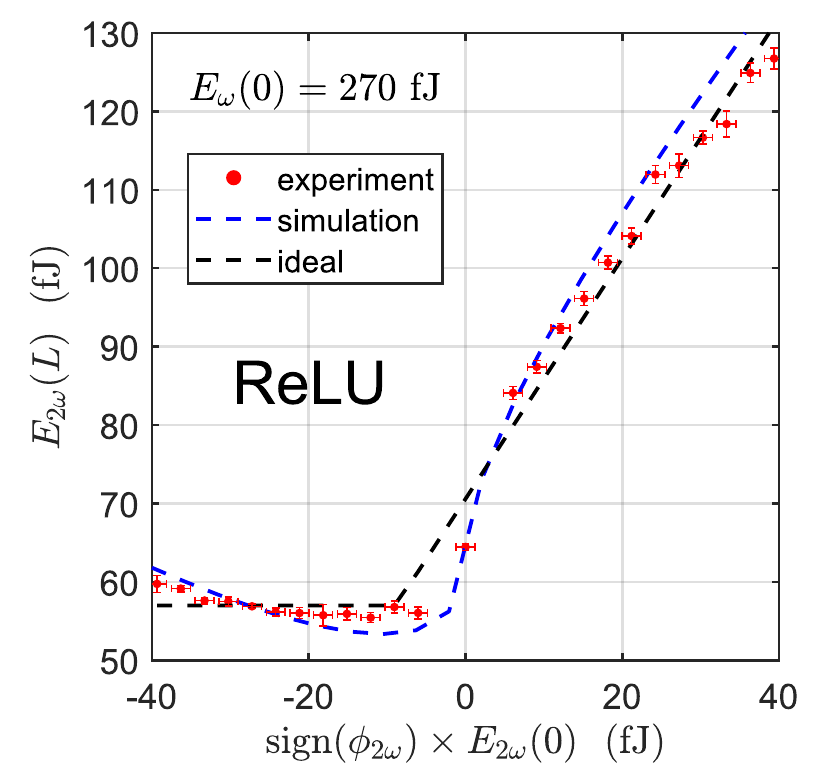}
\caption{Output signal pulse energy versus input signal pulse energy for both negative and positive inputs. There is good agreement between the ideal ReLU function (dashed black line), simulation (dashed blue line) and experimental results (red circles) for a bias pulse energy of $E_{\omega}(0)=270~\mathrm{fJ}$, and signal pulse energies of femtojoules per activation.}
\label{fig:fig3}
\end{figure}

The measured response of the all-optical ReLU is shown in Fig.~\ref{fig:fig3}. The nonlinear function given by the PPLN was measured using a free-space chip characterization setup. The source at $1045~\mathrm{nm}$ (signal) was a Yb:fiber mode-locked laser producing $75$-fs long pulses at a $250$-MHz repetition rate (Menlo Systems Orange). The same laser pumped a homemade degenerate optical parametric oscillator to generate the pulses at $2090~\mathrm{nm}$ (bias). The $2\omega$ and $\omega$ pulses were coupled into and out of the PPLN using reflective objectives focused on the waveguide facets. Finally, the relative phase of the $2\omega$ signal and $\omega$ bias was set using a delay arm, and the power varied using a tunable attenuator. See Supplement 1 for further details about the experimental setup.

Our experimental results show good agreement with the ideal ReLU function ($R^{2}=0.9895$), and demonstrates energy-efficient signal pulse energies in the regime of femtojoules per activation. Note that the important feature of the function is its nonlinear shape, since scaling/shifting the horizontal/vertical directions can be accomplished with linear optical transformations. In theory, the ideal ReLU function requires an arbitrarily long PPLN and low bias pulse energy. However, in practice we must choose the bias pulse energy so as to best approximate the ReLU function given our fixed device length. Thus, there are small discrepancies around $E_{2\omega}(0)=0$, since neither the SHG nor DOPA processes sufficiently saturate at the ultra-low energies. The maximum cutoff pulse energy is determined by the onset of supercontinuum generation from strong back-conversion processes, which undesirably degrades the pulse shape. To verify that the expected device response matches our physical picture of the operating principle, we also performed nonlinear pulse propagation simulations of the PPLN nanophotonic waveguide. See Supplement 1 for more details about the simulation methods.

Remarkably, we show that the PPLN nanophotonic waveguide can also approximate other commonly used variants of the ReLU function, simply by tuning the bias pulse energy. 
\begin{figure}[!h]
\centering
\includegraphics[scale=0.79]{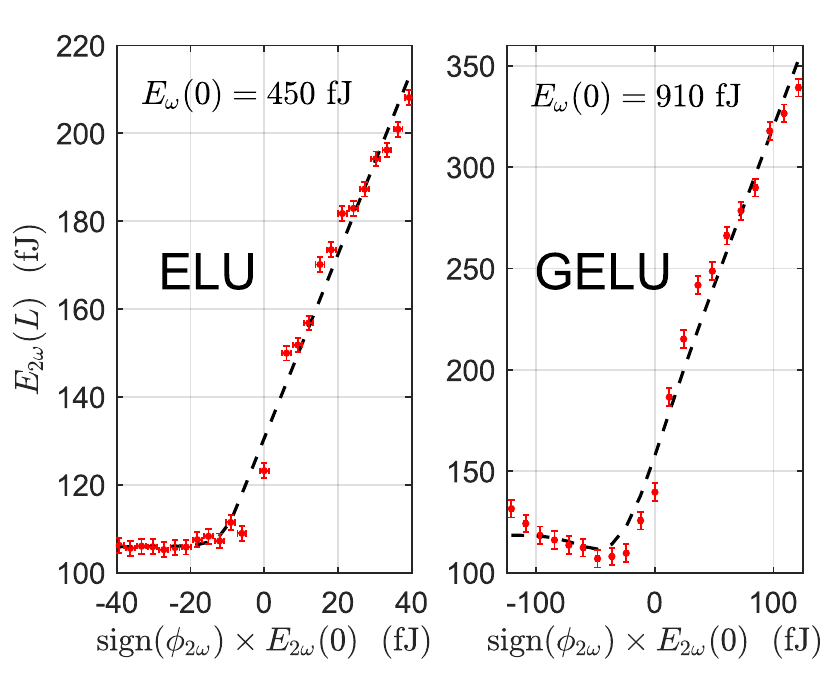}
\caption{Other variants of the ReLU function can be approximated by tuning the bias pulse energy. For example, the (a) ELU function using bias pulse energy of $E_{\omega}(0)=450~\mathrm{fJ}$ and (b) GELU function using bias pulse energy of $E_{\omega}(0)=910~\mathrm{fJ}$. Ideal function curves are shown by the dashed black lines, and experimental results with red circles.}
\label{fig:fig4}
\end{figure}
For example, the Exponential Linear Unit (ELU) defined as $\mathrm{ELU}(x)=x$ if $x>0$ and $\mathrm{ELU}(x)=\exp{(x)}-1$ if $x<0$, which has been shown to outperform the ReLU function in certain cases~\cite{clevert2015fast}, is achieved using a bias pulse energy of $E_{\omega}(0)=450~\mathrm{fJ}$ as shown in Fig.~\ref{fig:fig4}(a). In addition, we also implement the Gaussian Error Linear Unit (GELU) defined as $\mathrm{GELU}(x)=x\Phi(x)$ where $\Phi(x)$ is the Gaussian cumulative distribution using a bias pulse energy of $E_{\omega}(0)=910~\mathrm{fJ}$ as shown in Fig.~\ref{fig:fig4}(b). The GELU function is used extensively in Transformer networks for natural language processing, which are regularly amongst the largest deep learning models~\cite{brown2020language}. Thus, our all-optical PPLN nanophotonic waveguide implementation gains greater real-world applicability by being compatible with a wide range of existing deep learning models, especially the largest models where energy efficiency is paramount. Indeed, compatibility has been problematic in previous implementations of optical and optoelectronic nonlinear activation functions. By alleviating this problem, we expand the potential functionality of ONNs by avoiding the need to train new specialized models.

\subsection{Ultrafast time response}
Ideally, the time per activation should be near-instantaneous due to the ultrafast $\chi^{(2)}$ nonlinearity in lithium niobate. However, in practice, the response time is limited by the finite phase-matching bandwidth as well as non-zero group velocity mismatch, group velocity dispersion, and higher-order dispersion terms. To determine the response time of the device, we used the pump-probe technique commonly used to characterize all-optical switches~\cite{guo2021femtojoule,ono2020ultrafast,grinblat2019ultrafast} (see Supplement 1 for more details). In this case, the pump pulse is the $\omega$ pulse and the probe pulse is the $2\omega$ pulse. We measured the ultrafast ReLU dynamics by varying the time delay between the $\omega$ and $2\omega$ pulses at a fixed pulse energy.
\begin{figure}[!h]
\centering
\includegraphics[scale=0.79]{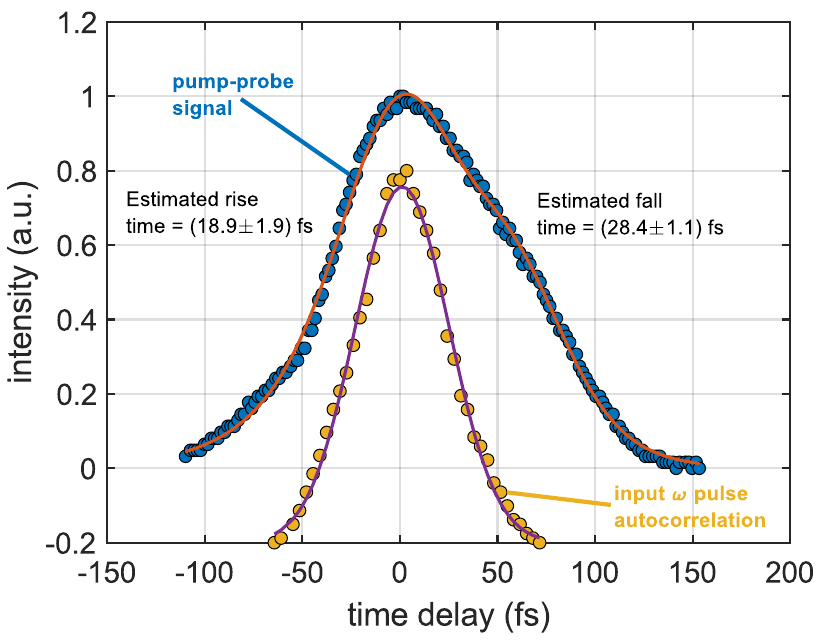}
\caption{Pump-probe ultrafast timing measurements of the ReLU dynamics. The autocorrelation (yellow circles shifted vertically for clarity) of the input $\omega$ pulse is well-explained by a Gaussian profile (purple line) with FWHM of $(56.4\pm1.5)~\mathrm{fs}$. The pump-probe signal obtained at a fixed pulse energy (blue circles) is fit (orange line) by convolving the input autocorrelation with exponential growth and decay for positive and negative time delays, respectively. The best fit yields a rise time of $(18.9\pm1.9)~\mathrm{fs}$ and a fall time of $(28.4\pm1.1)~\mathrm{fs}$.}
\label{fig:fig5}
\end{figure}
Fig.~\ref{fig:fig5} shows the intensity envelope of the pump-probe signal as the time delay is varied as well as the autocorrelation of the input $\omega$ pulse. The input autocorrelation is well-explained by a Gaussian profile with FWHM of $(56.4\pm1.5)~\mathrm{fs}$. We extract the characteristic rise and fall times by fitting the pump-probe signal with exponential growth and decay functions for positive and negative time delays, respectively, convolved with the input autocorrelation. The best fit yields a rise time of $(18.9\pm1.9)~\mathrm{fs}$ and a fall time of $(28.4\pm1.1)~\mathrm{fs}$. This implies that the characteristic response time of the ReLU dynamics is $(47.3\pm3.0)~\mathrm{fs}$, thus verifying that the ultrafast optical nonlinearity is responsible for the ReLU response, and ruling out the possibility of any slower optical nonlinearities such as photorefractive or thermo-optic effects. Therefore, we can reasonably regard the $2\omega$ signal pulse length of $\sim75~\mathrm{fs}$ as the time per activation for the all-optical ReLU. We note that that better dispersion engineering can lead to even faster activation times.  

\subsection{Simulated deep learning performance}
One distinct advantage of our approach is that, unlike previous all-optical and optoelectronic nonlinear activation functions, it can faithfully reproduce the ideal ReLU function. Therefore, we can leverage the large number of existing pretrained deep learning models that use the ReLU function (or its variants) for nonlinear activations. Although ONNs have been demonstrated that accurately reproduce linear operations such as matrix multiplication and convolution, the use of atypical nonlinear activation functions in the optical domain has required the training of new custom deep learning models. To improve upon this, we simulated the performance of the all-optical ReLU function when used as part of a pretrained convolutional neural network (CNN) for the prototypical task of MNIST handwritten digits image classification~\cite{deng2012mnist}.
\begin{figure}[!h]
\centering
\includegraphics[scale=0.79]{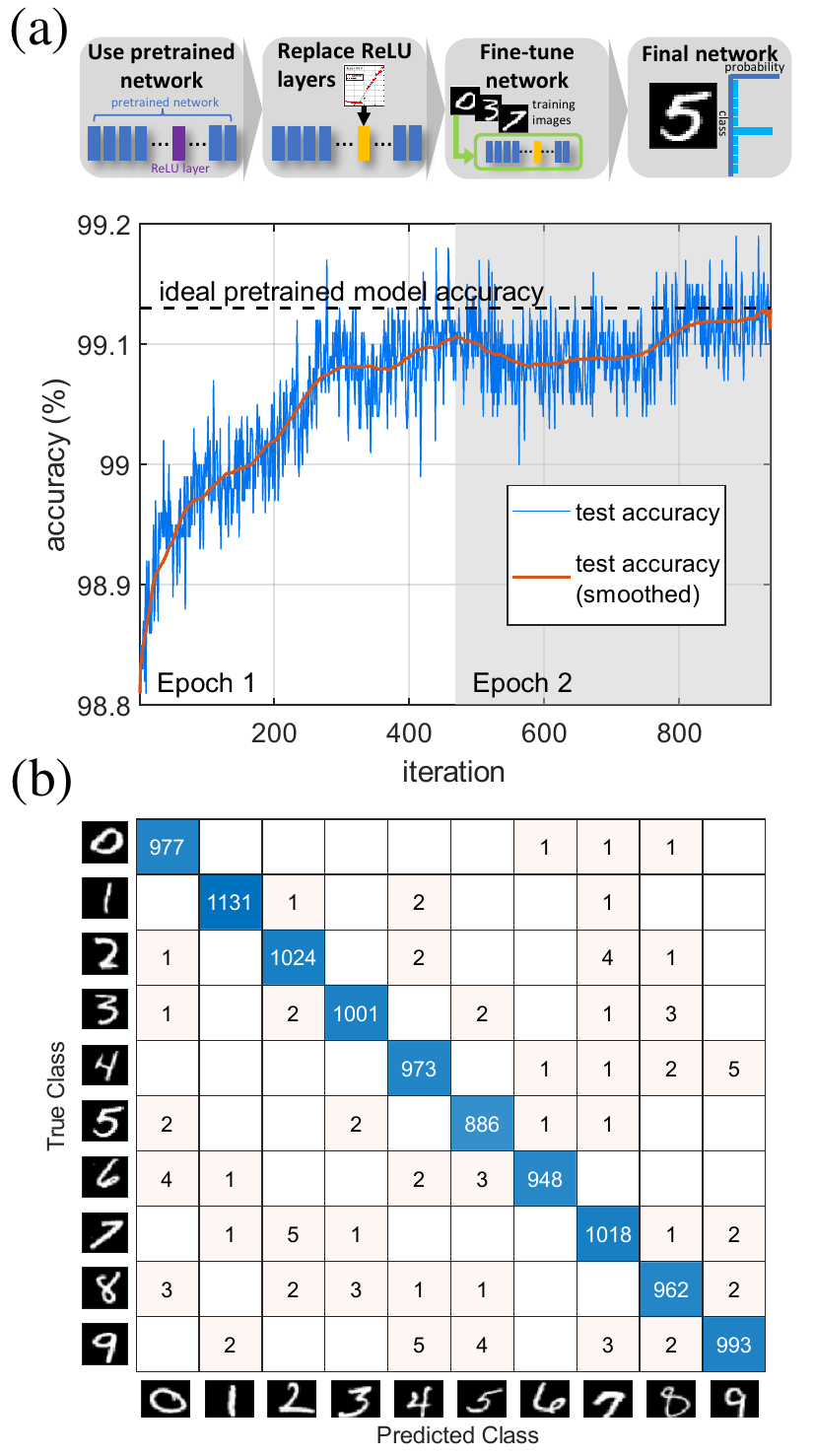}
\caption{Simulated deep learning performance of the experimentally measured all-optical ReLU function for MNIST handwritten digits image classification. (a) A pretrained CNN was used where the ideal ReLU layers are replaced with custom layers representing the experimentally measured ReLU response (after shifting/scaling) then fine-tuned by training for 2 epochs (batch size of 128) to improve the test accuracy (blue line) back to the ideal pretrained model accuracy (dashed black line). (b) Confusion matrix on the MNIST task for the final network, which achieved $99.13\%$ test accuracy.}
\label{fig:fig6}
\end{figure}
The MNIST dataset contains $28\times28$ pixels gray-scale images of handwritten digits with 50000 training samples and 10000 test samples. We used a standard CNN architecture (see Supplement 1 for full details) containing convolutional layers and ideal ReLU layers followed by a fully-connected layer and softmax classification output. The pretrained CNN achieved an ideal test accuracy of $99.13\%$. Next, the ideal ReLU layers were replaced with custom layers representing the experimentally measured ReLU response (after proper shifting/scaling) without changing any of the other layers. This caused a slight drop in test accuracy to $98.8\%$ due to the slight deviations between the experimentally measured and ideal ReLU functions. To remedy this, the CNN was then fine-tuned by training for only 2 epochs (the CNN sees each sample once per epoch) to regain the ideal pretrained model accuracy of $99.13\%$ as shown in Fig.~\ref{fig:fig6}. Fine-tuning is necessary for any analog hardware implementation due to unavoidable fabrication errors, noise and other non-idealities encountered~\cite{Bandyopadhyay:21}. Note that this method requires far less time compared to previous proposals for training new custom ONN models, which required $>25$ training epochs~\cite{guo2021backpropagation,williamson2019reprogrammable}. Therefore, our all-optical ReLU provides the missing link to allow ONNs to take advantage of existing pretrained models. We note that the softmax classification layer is yet to be faithfully implemented in an ONN which accounts for a small portion of the computation compared to the convolutions, matrix multiplications and ReLU nonlinear activations. 

\section{Discussion}
\subsection{Comparison of energy and time per activation}
\begin{figure}[!h]
\centering
\includegraphics[scale=0.79]{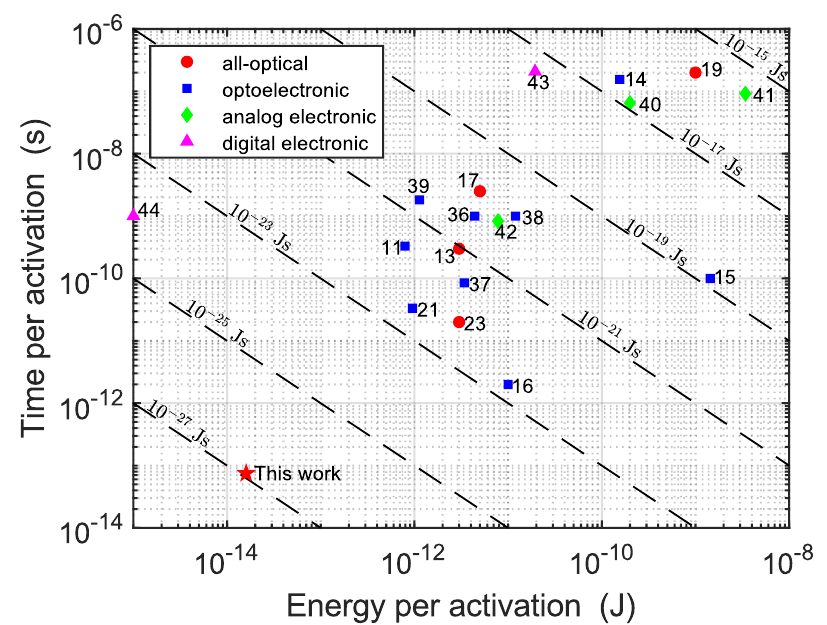}
\caption{Comparison of energy and time per activation of this work (red star) to previous all-optical (red circle), optoelectronic (blue square), analog electronic (green diamond), and digital electronic (magenta triangle) nonlinear activation functions. The numeric labels show reference numbers and dashed black lines show the energy-time product contours.}
\label{fig:fig7}
\end{figure}

In this section, we compare the PPLN nanophotonic waveguide to other optical~\cite{dejonckheere2014all,feldmann2019all,mourgias2019all,shi2021inp}, optoelectronic~\cite{porte2021complete,duport2012all,duport2016fully,shastri2016spike,ashtiani2021single,tait2017neuromorphic,crnjanski2021adaptive,amin2019ito,mesaritakis2016artificial}, analog electronic~\cite{oh2021energy,krestinskaya2018learning,huang2020analog}, and digital electronic~\cite{giordano2019analog} nonlinear activation functions to demonstrate the state-of-the-art performance of our device. In this case, the appropriate figure of merit is the energy-time product, which properly accounts for both the energy consumed and time taken per activation. To quantify the energy per activation, we follow the convention in \cite{williamson2019reprogrammable}, as being the energy needed to generate a $50\%$ change in the power transmission with respect to the transmission with null input. In this case, our device has an energy per activation of $\sim16~\mathrm{fJ}$. The bias pulse energy is not included since it is not dissipated and can be reused. The time per activation is given by the signal pulse width of $\sim75~\mathrm{fs}$, owing to the near-instantaneous $\chi^{(2)}$ nonlinearity of lithium niobate as explained in Section~3.2. Therefore, we achieve an energy-time product of $1.2\times10^{-27}~\mathrm{J\cdot s}$. The energy and time per activation of our device is compared to other experimental demonstrations in Fig.~\ref{fig:fig7}.
We attempted to consider device-level metrics wherever possible to provide a fair comparison, however, we acknowledge that this was not always possible for nonlinear activations as part of complete networks since fan-out and cascadability constraints impose additional energy and time costs. Despite this, the outstanding metrics of our device represents a significant breakthrough for optical nonlinear activation functions. For state-of-the-art digital electronics, such as the NVIDIA A100 Tensor Core GPU~\cite{choquette2021nvidia} based on 7-nm process node~\cite{xie2015performance}, we generously assume that the ReLU function consumes $\sim1~\mathrm{fJ}$ per activation, and occurs in a single $1~\mathrm{GHz}$ clock cycle. We see that, although our device still has an order of magnitude greater energy per activation, the time per activation is four orders of magnitude faster. Hence, we achieve an energy-time product that is three orders of magnitude better than state-of-the-art digital electronics. Our numerical simulations (see Supplement 1) predict that the PPLN nanophotonic waveguide can realistically achieve a ReLU-like response with sub-femtojoule energy per activation. This would even surpass the energy efficiency of state-of-the-art digital electronics. We attribute the discrepancy between our experimental results and the theoretically predicted limits for the energy scale to the imperfect phase-matching and fabrication error of our device. 

\subsection{Potential network architectures}
So far, we have demonstrated how PPLN nanophotonic waveguides can implement all-optical, ultrafast, energy-efficient nonlinear activation functions, which forms only one building block of a full neuron. In this section, we briefly discuss how our device can be integrated into a complete ONN architecture. In principle, the all-optical ReLU is compatible with most existing ONN architectures that can accurately implement linear operations such as matrix multiplication and convolutions. However, in practice, the speed bottleneck will likely be the encoding of information into the required coherent optical amplitudes. In this case, PPLN nanophotonic waveguides can be monolithically integrated with high-speed electro-optic modulators in thin-film lithium niobate, demonstrated to achieve bandwidths beyond $100~\mathrm{GHz}$~\cite{zhang2021integrated}. Furthermore, the light sources can also be integrated on-chip using thin-film lithium niobate optical parametric oscillators~\cite{lu2021ultralow}. Therefore, all the fundamental building blocks needed for a complete ONN in thin-film lithium niobate already exist. Given the rapid increases in scalability of thin-film lithium niobate photonics, we are confident that a complete ONN can be demonstrated in the near-future. One potential approach is to use Mach-Zehnder interferometer meshes~\cite{shen2017deep} or photonic tensor cores with waveguide cross-bar arrays~\cite{feldmann2021parallel} to implement the linear matrix multiplications, then cascaded into PPLN nanophotonic waveguides to perform nonlinear activations. Another promising method is to use a time-multiplexed architecture similar to ones demonstrated for coherent Ising machines~\cite{yamamoto2017coherent} or photonic reservoir computers~\cite{duport2016fully,duport2012all}. 

\section{Conclusion}

In conclusion, we have demonstrated an all-optical ultrafast ReLU function using a PPLN nanophotonic waveguide. It has an energy per activation of $\sim16~\mathrm{fJ}$ and time per activation of $\sim75~\mathrm{fs}$, thus achieving a state-of-the-art energy-time product of $1.2\times10^{-27}~\mathrm{J\cdot s}$. Furthermore, we demonstrated how the same device can be used to implement other common variants of the ReLU function, and showed how it can exploit existing pretrained deep learning models to greatly reduce training time. Given the simplicity of our device, and the rapid improvements in scalability of thin-film lithium niobate photonics, we envisage that it will be able to replace periphery digital electronic circuits for calculating nonlinear activations in ONNs. Therefore, we have presented a clear and practical path towards truly all-optical, energy-efficient photonic deep learning.

\section*{Funding} 
The authors gratefully acknowledge
support from ARO grant no. W911NF-18-1-0285, NSF
grant no. 1846273 and 1918549, AFOSR award FA9550-
20-1-0040, and NASA/JPL. The authors wish to
thank NTT Research for their financial and technical
support.

\section*{Acknowledgments}
The device nanofabrication was performed at the Kavli Nanoscience Institute (KNI) at Caltech.

\section*{Disclosures}
GHYL, RS, RN, RMG, AM: California Institute of Technology (P)

\section*{Data availability} Data underlying the results presented in this paper are not publicly available at this time but may be obtained from the authors upon reasonable request.

\section*{Supplemental document}
See Supplement 1 for supporting content. 

\bibliographystyle{ieeetr}
\bibliography{references}

\end{document}